\documentclass[journal,twoside,web]{ieeecolor} 

\usepackage{lcsys}
\usepackage{cite}
\usepackage{amsmath,amssymb,amsfonts}
\usepackage{graphicx}
\usepackage{textcomp}
\usepackage{adjustbox}
\usepackage{balance}
\usepackage{threeparttable}
\usepackage{soul}
\usepackage{verbatim}
\usepackage{url}
\usepackage{amssymb}
\usepackage{mathtools}
\usepackage{xparse}
\usepackage[percent]{overpic}
\usepackage{textcomp}
\usepackage{comment}
\usepackage{multirow}
\usepackage{array}
\usepackage{nicefrac} 
\usepackage{algorithm} 
\usepackage{algpseudocode} 
\usepackage{siunitx, booktabs}
\usepackage{newtxtext,newtxmath}
\usepackage{etoolbox}
\usepackage{hyperref}
\hypersetup{
    colorlinks=true,
    linkcolor=blue,      
    urlcolor=blue        
}

\newcommand{\bs}[1]{\boldsymbol{#1}}  
\newcommand{\ts}[1]{\text{#1}}

\makeatletter
\newcommand{\raisemath}[1]{\mathpalette{\raiseMath{#1}}}%
\newcommand{\raiseMath}[3]{\raisebox{#1}[0pt][0pt]{$#2#3$}}

\makeatother

\NewDocumentCommand{\qbar}{O{1.2pt} O{-6.8pt}}{
	\ensuremath{\mathrlap{\raisemath{#2}{\hspace*{#1}{\mathchar'26\mkern-9mu}}} q}%
}

\NewDocumentCommand{\qbarsmall}{O{1pt} O{-6.2pt}}{
	\ensuremath{\mathrlap{\raisemath{#2}{\hspace*{#1}{\mathchar'26\mkern-9mu}}} q}%
}

\NewDocumentCommand{\qbarssmall}{O{0.8pt} O{-5.0pt}}{
	\ensuremath{\mathrlap{\raisemath{#2}{\hspace*{#1}{\mathchar'26\mkern-9mu}}} q}%
}

\NewDocumentCommand{\qbars}{O{0.8pt} O{-5pt}}{
	\ensuremath{\mathrlap{\raisemath{#2}{\hspace*{#1}{\mathchar'26\mkern-9mu}}} q}%
}

\NewDocumentCommand{\qbarc}{O{0.5pt} O{-5.2pt}}{
	\ensuremath{\mathrlap{\raisemath{#2}{\hspace*{#1}{\mathchar'26\mkern-9mu}}} q}%
}

\NewDocumentCommand{\pbar}{O{0pt} O{-6.8pt}}{
	\ensuremath{\mathrlap{\raisemath{#2}{\hspace*{#1}{\mathchar'26\mkern-9mu}}} p}%
}

\renewcommand{\baselinestretch}{0.96}  

\def\BibTeX{{\rm B\kern-.05em{\sc i\kern-.025em b}\kern-.08em
    T\kern-.1667em\lower.7ex\hbox{E}\kern-.125emX}}
\makeatletter

\def\@oddfoot{\centering\footnotesize © 2026 IEEE. This work has been published in \emph{IEEE Control Systems Letters}. Personal use of this material is permitted. Permission from IEEE must be obtained for all other uses.\hfil}
\def\@evenfoot{\@oddfoot}
\makeatother

\begin{document}

\title{\vspace{0.5ex} A Class of Axis--Angle Attitude Control Laws\\for Rotational Systems}
\author{Francisco M. F. R. Gon\c{c}alves, \IEEEmembership{Graduate Student Member, IEEE}, Ryan M. Bena, \IEEEmembership{Member, IEEE}, \mbox{and N\'{e}stor O. P\'{e}rez-Arancibia, \IEEEmembership{Member, IEEE}}
\thanks{This work was supported in part by the Washington State University (WSU) Foundation and the Palouse Club through a Cougar Cage Award to N.\,O.\,P\'erez-Arancibia and in part by the WSU Voiland College of Engineering and Architecture through a start-up package to N.\,O.\,P\'erez-Arancibia.}
\thanks{F.\,M.\,F.\,R.\,Gon\c{c}alves and R.\,M.\,Bena contributed equally to this work.}
\thanks{F.\,M.\,F.\,R.\,Gon\c{c}alves and N.\,O.\,P\'erez-Arancibia are with the School of Mechanical and Materials Engineering, Washington State University (WSU), Pullman, WA 99164-2920, USA. R.\,M.\,Bena is with the Department of Mechanical and Civil Engineering, California Institute of Technology, Pasadena, CA 91125-2100, USA. Corresponding authors' email: {\tt francisco.goncalves@wsu.edu} (F.\,M.\,F.\,R.\,G.); {\tt n.perezarancibia@wsu.edu} (N.\,O.\,P.-A.).}
\vspace{-5.0ex}}

\maketitle

\begin{abstract}
We introduce a new class of attitude control laws for rotational systems; the proposed framework generalizes the use of the Euler \mbox{axis--angle} representation beyond quaternion-based formulations. Using basic Lyapunov stability theory and the notion of extended class $\mathcal{K}$ function, we developed a method for determining and enforcing the global asymptotic stability of the single fixed point of the resulting \mbox{\textit{closed-loop}} (CL) scheme. In contrast with traditional \mbox{quaternion-based} methods, the introduced generalized \mbox{axis--angle} approach enables greater flexibility in the design of the control law, which is of great utility when employed in combination with a switching scheme whose transition state depends on the angular velocity of the controlled rotational system. Through simulation and \mbox{real-time} experimental results, we demonstrate the effectiveness of the developed formulation. According to the recorded data, 
in the execution of \mbox{high-speed} \mbox{tumble-recovery} maneuvers, the new method consistently achieves shorter stabilization times and requires lower control effort relative to those corresponding to the \mbox{quaternion-based} and 
\mbox{geometric-control} methods used as benchmarks.
\end{abstract}


\section{Introduction}
\label{sec:introduction}

\IEEEPARstart{N}{umerous} different types of attitude controllers have been proposed for stabilizing and commanding the trajectory of rotational systems that can be modeled as rigid bodies\cite{GoncalvesFMFR2024I,GoncalvesFMFR2024III,SchlanbuschD2012,BenaRM2022,BenaRM2023I,MayhewCG2009,MayhewCG2011I,MayhewCG2011II, PratamaB2018,MokhtariA2004,KangCW2011,LeeT2011,LeeT2018, WuG2014,WeiJ2017,ThunbergJ2014}. These controllers can be classified into three main types: (i)~quaternion-based, (ii)~Euler-angles--based, \mbox{and (iii)} rotation-matrix--based~(geometric). The \mbox{limitations of} attitude controllers based on Euler angles have been extensively studied and discussed in the technical literature---for example, see\cite{ChaturvediNA2011} and references therein. Due to these limitations, \mbox{quaternion-based} and geometric methods have been the preferred choices for representing and implementing robust \mbox{high-performance} attitude controllers. Although these two formulations implicitly contain the knowledge about the Euler axis and associated rotation angle as defined in\cite{KuipersQuaternions}, this information is generally not directly used in controller design. 

For example, \mbox{quaternion-based} \mbox{time-invariant} attitude control laws usually include a proportional term that is formed by scaling the \mbox{attitude-error} Euler axis with $\sin\frac{\Theta_{\ts{e}}}{2}$, where $\Theta_{\ts{e}}$ is the corresponding instantaneous rotation error. However, there is no reason for this type of formulation other than ensuring continuity in the entire rotational space and simplifying the analysis of the resulting \textit{closed-loop} (CL) system. Counterintuitively, however, in most practical applications, this continuous formulation is modified to obtain a switching or hybrid scheme in order to avoid unwinding behavior\mbox{\cite{BenaRM2022,BenaRM2023I,MayhewCG2009,MayhewCG2011I,MayhewCG2011II}}. Another characteristic of continuous \mbox{quaternion-based} control laws is that, for rotational errors larger than \mbox{$\pi$\,rad}, the proportional control effort decreases as the rotational error about the \mbox{attitude-error} Euler axis increases. Also, it can be shown that \mbox{geometric-based} attitude control laws implicitly apply torques in the direction of the shorter rotational trajectory required to eliminate the attitude error. As discussed in\cite{GoncalvesFMFR2024I}~and~\cite{GoncalvesFMFR2024III}, this direction choice is not necessarily the best for every kinematic situation, depending on the specific performance objectives of the problem of interest.

In this letter, we introduce a class of attitude control laws applicable to a wide gamut of rigid rotational systems moving in the \mbox{\textit{three-dimensional}~($3$D)} space, including \textit{uncrewed aerial vehicles} (UAVs), satellites, and microrobotic swimmers. The proposed method generalizes the use of the Euler \mbox{axis--angle} representation in the formulation of the feedback law that stabilizes and drives the rotational motion of the controlled system. It is well known that global asymptotic stability on the special orthogonal group in three dimensions, $\bs{\mathcal{SO}}(3)$, cannot be achieved with the use of continuous \mbox{time-invariant} feedback controllers. A common technique used to overcome this topological obstruction is to design discontinuous control laws\cite{MayhewCG2011II}; another option is to define the CL dynamics of the controlled system on a covering that generates a unique coordinate representation of $\bs{\mathcal{SO}}(3)$, rather than on $\bs{\mathcal{SO}}(3)$ itself. Following this latter approach, by combining Lyapunov stability theory and the notion of extended class $\mathcal{K}$ function, we developed a method that ensures the existence of a unique CL fixed \textit{\mbox{attitude-error} quaternion} (AEQ) and the global asymptotic stability of the corresponding equilibrium point, which depends on \mbox{positive-definiteness} conditions that the controller gains must satisfy. 

A main contribution of the work presented in this letter is the design flexibility enabled by a generalized formulation of the proportional term in the proposed control law; specifically, it allows the user to select a suitable proportional control function based on the platform \mbox{and/or} application, with stability guarantees. Furthermore, in contrast to existing \mbox{quaternion-based} control methods, the new generalized \mbox{axis--angle} attitude control approach ensures a greater proportional action the farther the system's state is from the stable CL fixed AEQ. This characteristic is especially useful when a scheme of the proposed type is employed in combination with intelligent switching methods capable of taking into account the angular velocity when selecting the direction in which the proportional component of the input torque is applied during operation, as for example done in the case presented in\cite{GoncalvesFMFR2024I}. We tested the functionality and performance of the introduced approach through simulations and outdoor flight tests. The obtained simulation and experimental data show that, compared to two other \mbox{high-performance} benchmark controllers---one \mbox{quaternion-based} and another geometric---a scheme of the new type is unequivocally superior. Specifically, measurements obtained through dozens of \mbox{high-speed} \mbox{tumble-recovery} maneuvers show that the presented approach consistently achieves shorter stabilization times and requires less control effort, from a statistical standpoint. 

\vspace{1ex}
\textit{\textbf{Notation---}}
\begin{enumerate}
\item Lowercase symbols represent scalars, e.g., $p$; bold lowercase symbols denote vectors, e.g., $\bs{p}$; bold uppercase symbols denote matrices, e.g., $\bs{P}$; and, bold crossed lowercase symbols denote quaternions, \mbox{e.g., $\bs{\pbar}$}.
\item The set of unit $3$D vectors is denoted by $\bs{\mathcal{S}}^2$.
\item The sets of reals and positive reals are denoted by $\mathbb{R}$ and $\mathbb{R}_{> 0}$, respectively. The set of integers is denoted by $\mathbb{Z}$.
\item The symbols $\times$ and $\otimes$ denote the \mbox{cross-product} of two vectors and multiplication between two quaternions, respectively.
\item Throughout the rest of the paper, we use the notion of extended class $\mathcal{K}$ function as defined in\cite{AmesAD2017}.
\end{enumerate}

\section{Background}
\label{Section02}
\begin{figure}[t!]
\vspace{1.4ex}
\begin{center}
\includegraphics{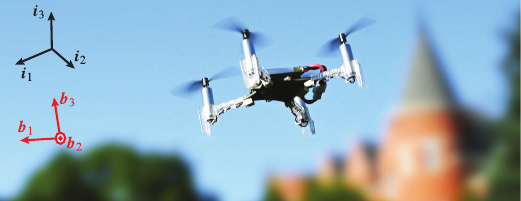}
\end{center}
\vspace{-2ex}
\caption{\mbox{\textbf{Flight control tests and frames of reference.}} This picture shows the UAV, a \mbox{Crazyflie\,2.1}, used in the control experiments, and the inertial and \mbox{body-fixed} frames of reference, \mbox{$\bs{\mathcal{I}} = \left\{\bs{i}_1, \bs{i}_2, \bs{i}_3\right\}$} and \mbox{$\bs{\mathcal{B}} = \left\{\bs{b}_1, \bs{b}_2, \bs{b}_3\right\}$}. Here, $\bs{\mathcal{B}}$, whose origin coincides with the robot's CoM, is shown shifted for clarity. \label{Fig01}}
\vspace{-2ex}
\end{figure}

We first describe the dynamics of a rigid body rotating in the $3$D space. As shown in \mbox{Fig.\,\ref{Fig01}}, \mbox{$\bs{\mathcal{B}}=\left\{\bs{b}_1, \bs{b}_2, \bs{b}_3\right\}$} denotes the \mbox{body-fixed} frame of reference, with its origin coinciding with the body's \textit{center of mass} (CoM), used for kinematic description and dynamic analysis. Consistent with this definition, \mbox{$\bs{\mathcal{I}}=\left\{\bs{i}_1, \bs{i}_2, \bs{i}_3\right\}$} denotes the chosen inertial frame of reference fixed to the planet. As discussed in\cite{BenaRM2022} and \cite{BenaRM2023I}, using quaternions and Euler's second law, the \mbox{open-loop} rotational dynamics of the system can be described as
\begin{subequations}
\begin{align}
\begin{split}
\bs{\dot{\qbar}} &= \frac{1}{2} \bs{\qbar}\otimes
\begin{bmatrix}
0 \\
\bs{\omega} 
\end{bmatrix}
\end{split},
\label{EQN01a} \\
\begin{split}
\bs{\dot{\omega}} &=  \bs{J}^{-1}\left(\bs{\tau}-\bs{\omega}\times \bs{J}\bs{\omega}\right),
\label{EQN01b}
\end{split}
\end{align} 
\end{subequations}
in which $\bs{\qbar}$ is a unit quaternion that represents the orientation of $\bs{\mathcal{B}}$ relative to $\bs{\mathcal{I}}$; $\bs{\omega}$ is the angular velocity of $\bs{\mathcal{B}}$ with respect to $\bs{\mathcal{I}}$, written in $\bs{\mathcal{B}}$ coordinates; $\bs{J}$ is the inertia matrix of the body, written in $\bs{\mathcal{B}}$ coordinates; and, $\bs{\tau}$ is the torque applied to the system written in $\bs{\mathcal{B}}$ coordinates and, in closed loop, generated by a control law. 

As specified by Euler's rotation theorem, any sequence of rotations of a rigid body in the $3$D space is equivalent to a single rotation of amount $\Theta$ about an axis $\bs{u}$ that passes through the body's CoM. This information can be stored in the form of a unit quaternion as \mbox{$\bs{\qbar} = \left[m\,\,\bs{n}^T\right]^T$}, where \mbox{$m = \cos{\frac{\Theta}{2}}$} and \mbox{$\bs{n} = \bs{u}\sin{\frac{\Theta}{2}}$}. In the case specified by (\ref{EQN01a}), $\Theta$ is the amount that the modeled body must be rotated about $\bs{u}$ to reach the attitude of $\bs{\mathcal{B}}$ starting from that of $\bs{\mathcal{I}}$. Accordingly, the instantaneous AEQ is given by \mbox{$\bs{\qbar}_\text{e} = \bs{\qbar}^{-1} \otimes \bs{\qbar}_\text{d} = \left[m_{\ts{e}}~\bs{n}_{\ts{e}}^T\right]^T$}, with \mbox{$m_{\ts{e}} = \cos{\frac{\Theta_{\ts{e}}}{2}}$} and \mbox{$\bs{n}_{\ts{e}} = \bs{u}_{\ts{e}}\sin{\frac{\Theta_{\ts{e}}}{2}}$}, where $\Theta_{\ts{e}}$ is the amount that $\bs{\mathcal{B}}$ must be rotated about the \mbox{attitude-error} \mbox{Euler axis, $\bs{u}_{\ts{e}}$,} to reach the orientation of the desired \mbox{body-fixed} frame, $\bs{\mathcal{B}}_\ts{d}$; and, $\bs{\qbar}_\text{d}$ is a unit quaternion that represents the \mbox{attitude of $\bs{\mathcal{B}}_\ts{d}$} relative to $\bs{\mathcal{I}}$. In practice, $\bs{\qbar}_\text{d}$ is either defined by the user, as described in \mbox{Section\,\ref{SEC04}}, or generated by a position control scheme, as explained in\cite{BenaRM2022}. Any \mbox{real-time} control algorithm based on the formulation specified by \eqref{EQN01b} requires the desired angular velocity of the controlled rotational system in $\bs{\mathcal{B}}$ coordinates, $\bs{\omega}_{\ts{d}}$. To compute this variable, we first obtain the corresponding desired \mbox{angular-velocity} quaternion, written in $\bs{\mathcal{B}}_\ts{d}$ coordinates, according to
\begin{align}
\begin{bmatrix}
0 \\
\bs{\hat{\omega}}_\text{d}
\end{bmatrix}
= 2\bs{\qbar}_\ts{d}^{-1} \otimes \bs{\dot{\qbar}}_\text{d}.
\label{EQ02}
\end{align}
Then, we compute \mbox{$\bs{\omega}_\text{d} = \bs{S}^T \bs{S}_\ts{d} \bs{\hat{\omega}}_\text{d}$}, where $\bs{S}_\ts{d}$ transforms vectors from $\bs{\mathcal{B}}_\ts{d}$ to $\bs{\mathcal{I}}$ coordinates, and $\bs{S}^T$ transforms vectors from $\bs{\mathcal{I}}$ to $\bs{\mathcal{B}}$ coordinates. Thus, using these kinematic variables, we can define a control law that computes the torque input as
\begin{align}
\boldsymbol{\tau}_{\text{b}} = \boldsymbol{J}\left(k_{\boldsymbol{\qbars}} \boldsymbol{n}_{\text{e}} + k_{\boldsymbol{\omega}} \boldsymbol{\omega}_{\text{e}} + \boldsymbol{\dot{\omega}}_{\text{d}}\right) + \boldsymbol{\omega}\times\boldsymbol{J}\boldsymbol{\omega},
\label{EQN03}
\end{align}
where $k_{\bs{\qbars}}$ and $k_{\bs{\omega}}$ are scalar positive real controller gains; \mbox{$\bs{\omega}_{\ts{e}} = \bs{\omega}_{\ts{d}} - \bs{\omega}$} is the \mbox{angular-velocity} tracking error; $\bs{J}\bs{\dot{\omega}}_{\ts{d}}$ is a feedforward term that approximately cancels the \mbox{left-hand} side of~\eqref{EQN01b} and is included with the objective of providing faster tracking performance; and, $\bs{\omega}\times\bs{J}\bs{\omega}$ is a \mbox{feedback-linearization} term that cancels the nonlinearity present in~\eqref{EQN01b}.

As discussed in~\cite{BenaRM2022}, the CL rotational dynamics obtained by substituting the \mbox{right-hand} side of (\ref{EQN03}) into \eqref{EQN01b} exhibit two equilibria corresponding to the same kinematic condition but with different stability properties---one asymptotically stable and the other unstable. It is straightforward to see that, at both fixed points, \mbox{$\bs{n}_{\ts{e}} = \bs{0}$}. Therefore, if the state of the CL system were to be exactly at the unstable equilibrium, the control torque specified by (\ref{EQN03}) would not compel the controlled rotational system to execute a \mbox{$2\pi$-rad} rotation and converge to the stable equilibrium, thus preventing global asymptotic stability on $\bs{\mathcal{SO}}(3)$. However, in any \mbox{real-world} application, any deviation from the unstable fixed point would make \mbox{$\bs{n}_{\ts{e}} \neq \bs{0}$} and, therefore, for all practical purposes, we can confidently assume that the torque generated according to (\ref{EQN03}) forces the CL system's state to converge to the stable equilibrium. Additionally, note that since $\bs{n}_{\ts{e}}$ is scaled by the factor $\sin\frac{\Theta_{\ts{e}}}{2}$, the first term in (\ref{EQN03}) does not provide the maximum allowable control effort when the controlled rotational system is at the farthest orientation from the stable CL equilibrium AEQ. In fact, this term reaches its minimum value at the unstable CL fixed AEQ, where \mbox{$\Theta_\text{e}=2\pi$\,rad}; its maximum at \mbox{$\Theta_\text{e}=\pi$\,rad}; and, its minimum again at the stable CL equilibrium AEQ, where \mbox{$\Theta_\text{e}=0$\,rad}. These observations contradict the expected behavior of the proportional action of a feedback controller. We address these representational, stability, and performance issues in the next section by proposing a new generalized \mbox{axis--angle} attitude law for controlling rotational systems in the $3$D space.

\section{Generalized \mbox{Axis--Angle} Attitude Control}
\label{SEC03}
\subsection{A New Class of Control Laws}
\label{SEC03A}
Let \mbox{$\gamma\hspace{-0.4ex}: \mathbb{R}\mapsto\mathbb{R}$} be an extended class $\mathcal{K}$ function. Then, for a system whose dynamics evolve on $\bs{\mathcal{SO}}(3)$, the rotational errors to be minimized can be represented as
\begin{align}
\bs{\alpha}_\ts{e} &= \gamma(\Theta_\ts{e})\bs{u}_\ts{e}\quad\ts{and}\quad \bs{\omega}_{\ts{e}},
\label{EQN04}
\end{align}
where $\bs{\alpha}_\ts{e}$ is a \textit{scaled Euler axis} (SEA) aligned with $\bs{u}_\ts{e}$, whose magnitude---measured by any vector norm---scales with the value of $\Theta_\ts{e}$. Without loss of coverage of $\bs{\mathcal{SO}}(3)$, $\Theta_\ts{e}$ can be restricted to any \mbox{half-open} interval spanning $2\pi$ that includes $0$; in this letter, we consider \mbox{$\Theta_{\ts{e}} \in [0,2\pi)$}. Consistent with this selection, for any \mbox{$\Theta_\ts{e} \neq 0$}, an \mbox{axis--angle} pair, $\mbox{$\{\bs{u}_\ts{e},\Theta_\ts{e}\}$}$, uniquely defines one of the two possible rotations---with opposite directions---that would cancel the instantaneous attitude error of the controlled rotational system. When \mbox{$\Theta_\ts{e} = 0$}, the corresponding rotation axis is not uniquely defined and can take any value in $\bs{\mathcal{S}}^2$; however, $\bs{\alpha}_\ts{e}$ is still well defined. To initialize simulations and experiments, we select the direction of rotation at initial time $t_0$ that minimizes, assuming \mbox{$\Theta_{\ts{e}}\in\left[0,2\pi\right)$}, a \mbox{two-valued} cost function, as described in\cite{GoncalvesFMFR2024I}. Thus, using the errors specified by (\ref{EQN04}), we define the proposed generic control law as
\begin{align}
\bs{\tau}_{\gamma} = \bs{J}\left(k_{\bs{\alpha}}\bs{\alpha}_\ts{e} + k_{\delta}\bs{\dot{\alpha}}_\ts{e} + k_{\bs{\omega}} \bs{\omega}_\ts{e} + \bs{\dot{\omega}}_\ts{d}\right) + \bs{\omega}\times\bs{J}\bs{\omega},
\label{EQN05}
\end{align}
in which \mbox{$k_{\bs{\alpha}}$, $k_{\delta}$, and $k_{\bs{\omega}}\in\mathbb{R}_{>0}$}. Then, from simple substitution of (\ref{EQN04}) into (\ref{EQN05}), we obtain that 
\begin{align}
\begin{split}
\bs{\tau}_{\gamma} =& ~\bs{J}\left[k_{\bs{\alpha}}\gamma(\Theta_\ts{e})\bs{u}_\ts{e} + k_{\delta} \gamma(\Theta_\ts{e})\bs{\dot{u}}_\ts{e} + k_{\delta} \dot{\gamma}(\Theta_\ts{e})\bs{u}_\ts{e} \right. \\
&+ \left. k_{\bs{\omega}}\bs{\omega}_\ts{e} + \dot{\bs{\omega}}_\ts{d}\right] + \bs{\omega}\times\bs{J}\bs{\omega}.
\label{EQN06}
\end{split}
\end{align}

\subsection{Closed-Loop Dynamics and Equilibrium Points}
\label{SEC03B}
It is straightforward to see that for the control input defined by (\ref{EQN06}), the CL \mbox{rotational-error} dynamics can be written as 
\begin{subequations}
\begin{align}
\bs{\dot{\qbar}}_{\ts{e}} &= \frac{1}{2}
\begin{bmatrix}
0 \\
\bs{\omega}_{\ts{e}}
\end{bmatrix}
\otimes \bs{\qbar}_{\ts{e}},\label{EQN07a}\\
\begin{split}
\bs{\dot{\omega}}_\ts{e} &= -k_{\bs{\alpha}}\gamma(\Theta_\ts{e})\bs{u}_\ts{e} - k_{\delta} \left[ \gamma(\Theta_\ts{e})\bs{\dot{u}}_\ts{e}+\frac{d\gamma}{d\Theta_\ts{e}}(\bs{u}_\ts{e}^T\bs{\omega}_\ts{e})\bs{u}_\ts{e}\right]\\
&~~~-k_{\bs{\omega}}\bs{\omega}_\ts{e},
\end{split}
\label{EQN07b}
\end{align}
\end{subequations}
in whose derivation we used that \mbox{$\dot{\Theta}_\ts{e} = \bs{u}_\ts{e}^T\bs{\omega}_\ts{e}$}. Next, to find the equilibrium point(s) of the system given by \mbox{(\ref{EQN07a})--(\ref{EQN07b})}, we solve the set of algebraic equations specified by 
\begin{subequations}
\begin{align}
\begin{split}
-\frac{1}{2}\bs{n}_{\ts{e}}^T\bs{\omega}_{\ts{e}}  &= 0,
\label{EQN08a}
\end{split}\\
\begin{split}
- \frac{1}{2}\left( \bs{n}_{\ts{e}} \times \bs{\omega}_{\ts{e}}  - m_{\ts{e}}\bs{\omega}_{\ts{e}}   \right) &= \bs{0}_{3\times1},\label{EQN08b} \end{split}\\ 
\begin{split}
-k_{\bs{\alpha}}\gamma(\Theta_\ts{e})\bs{u}_\ts{e} - k_{\delta}\gamma(\Theta_\ts{e})\dot{\bs{u}}_\ts{e} - k_{\bs{\omega}}\bs{\omega}_\ts{e} &= \bs{0}_{3\times1}. \label{EQN08c} \end{split}
\end{align}
\end{subequations}
For (\ref{EQN08b}) to be satisfied, both terms inside the parenthesis must be zero because they are orthogonal. For both these terms to be zero, one of the following statements must hold:~\mbox{(i)~$\bs{n}_{\ts{e}}\parallel \bs{\omega}_{\ts{e}}$ and $m_\ts{e}=0$};~\mbox{(ii)~$\bs{\omega}_{\ts{e}} = \bs{0}$};~\mbox{(iii)~$\bs{n}_{\ts{e}} = \bs{\omega}_{\ts{e}} = \bs{0}$}. Also, for (\ref{EQN08a}) to be satisfied, either (ii) or (iii) is true, \mbox{$\bs{n}_\ts{e}=\bs{0}$}, or \mbox{$\bs{n}_{\ts{e}}\perp\bs{\omega}_{\ts{e}}$}. Since $\bs{n}_{\ts{e}}$ and $\bs{\omega}_{\ts{e}}$ cannot simultaneously be orthogonal and parallel, the only viable option is either (ii) or (iii). The fulfillment of either (ii) or (iii) requires that the solution to \mbox{(\ref{EQN08a})--(\ref{EQN08c})} satisfies \mbox{$\bs{\omega}^{\star}_{\ts{e}} = \bs{0}$}. Since $\bs{u}_\ts{e}$ is a vector evolving on $\bs{\mathcal{S}}^2$, it follows that \mbox{$\bs{u}_\ts{e}\perp\bs{\dot{u}}_\ts{e}$} and, therefore, for (\ref{EQN08c}) to be satisfied, \mbox{$\gamma(\Theta_{\ts{e}}) = 0$}. Last, recalling that $\gamma(\Theta_\ts{e})$ is an extended class $\mathcal{K}$ function and, therefore, \mbox{$\gamma(0) = 0$}, we conclude that \mbox{$\Theta_\ts{e}^{\star}=0$}. 

For the law specified by (\ref{EQN06}), we selected $\gamma(\Theta_\ts{e})$ to be an extended class $\mathcal{K}$ function in order to enforce the existence of a unique zero for $\gamma(\Theta_{\ts{e}})$ and a unique equilibrium point for the CL dynamics given by \mbox{(\ref{EQN07a})--(\ref{EQN07b})}, over the selected range of $\Theta_{\ts{e}}$. In summary, the CL dynamics resulting from using the control input specified by (\ref{EQN06}) \mbox{exhibit a unique equilibrium,} given by the \mbox{quaternion--vector} pair \mbox{$\bs{\qbar}^{\star}_{\ts{e}} = \left[1~0~0~0\right]^T$} and \mbox{$\bs{\omega}^{\star}_{\ts{e}} = \bs{0}$}, for \mbox{$\Theta_{\ts{e}} \in \left[0,2\pi\right)$}. This result does not contradict the topological obstruction discussed in\cite{ChaturvediNA2011}---according to which all continuous \mbox{time-invariant} CL vector fields on $\bs{\mathcal{SO}}(3)$ must exhibit multiple equilibria---as the CL dynamics are defined on an \mbox{axis--angle} covering that generates a unique coordinate representation of each rotation in $\bs{\mathcal{SO}}(3)$, rather than on $\bs{\mathcal{SO}}(3)$ itself. Furthermore, for \mbox{$\Theta_{\ts{e}} \in \left(-\infty,\infty \right)$}, the term $\gamma(\Theta_{\ts{e}})\bs{u}_{\ts{e}}$ is discontinuous at \mbox{$\Theta_{\ts{e}} = 2{\pi}k$\,rad}, for all nonzero \mbox{$k \in \mathbb{Z}$}. As a result, the CL dynamics specified by \mbox{(\ref{EQN07a})--(\ref{EQN07b})} are also discontinuous, which forces the existence of a unique fixed point, similarly to the cases presented in\cite{MayhewCG2011II}. However, discontinuities produce numerous experimental and theoretical problems; for example, to analyze the differential equations of the CL system specified by \mbox{(\ref{EQN07a})--(\ref{EQN07b})} when \mbox{$\Theta_{\ts{e}} \in \left(-\infty,\infty \right)$}, we must use Carath\'{e}odory's notion of solution, as stated in \mbox{[\citenum{HajekO1979},~Definition\,2.1]}. For these reasons, \mbox{$\Theta_{\ts{e}} \in \left[0,2\pi\right)$} is a superior choice for the problem considered in this letter. 

\subsection{Stability Analysis}
\label{SEC03C}
\noindent\makebox[3.48in][s]{\textbf{Theorem\,1.} Let the attitude and \mbox{angular-velocity} refer-}\newline ences, $\bs{\qbar}_\ts{d}$ and $\bs{\omega}_\ts{d}$, be smooth and bounded functions of time. Also, let the real controller gains satisfy \mbox{$k_{\delta}>0$}, \mbox{$k_{\bs{\omega}}>0$}, and \mbox{$k_{\bs{\alpha}}>\frac{k_{\delta}k_{\bs{\omega}}}{4}$}. Then, the fixed point \mbox{$\{\Theta_\ts{e}^{\star},\bs{\omega}_\ts{e}^{\star}\}$} of the CL \mbox{state-space} rotational dynamics specified by \eqref{EQN07a} and \eqref{EQN07b} is globally asymptotically stable on the domain 
\mbox{$\bs{\mathcal{D}}
=\left\{\,\{\Theta_{\ts{e}},\bs{\omega}_{\text{e}}\}\;\middle|\;
\Theta_{\text{e}}\in[0,2\pi),\ \bs{\omega}_{\text{e}}\in\mathbb{R}^3\,\right\}$}, for any initial state $\{\Theta_{\ts{e}}(t_0),\bs{\omega}_{\ts{e}}(t_0)\}\in\bs{\mathcal{D}}$ and \mbox{$\bs{u}_{\ts{e}}(t_0) \in \bs{\mathcal{S}}^2$}.
\vspace{1ex}

\noindent\textit{Proof.} Let a candidate Lyapunov function be
\begin{align}
\begin{split}
V =& 
\begin{bmatrix}
\gamma(\Theta_\ts{e})\bs{u}^T_\ts{e} & \bs{\omega}^T_\ts{e}\\
\end{bmatrix}
\begin{bmatrix}
\dfrac{k_{\delta}^2}{2k_{\bs{\alpha}}} & \dfrac{k_{\delta}}{2k_{\bs{\alpha}}}
\vspace{1ex}
\\
\dfrac{k_{\delta}}{2k_{\bs{\alpha}}} & \dfrac{1}{2k_{\bs{\alpha}}}
\end{bmatrix}
\begin{bmatrix}
\gamma(\Theta_\ts{e})\bs{u}_\ts{e} \\ \bs{\omega}_\ts{e}\\
\end{bmatrix}
\\ &\hspace{0.2ex}+ \int_0^{\Theta_\ts{e}}\gamma(\phi)d\phi,
\end{split}
\label{EQN09}
\end{align}
for \mbox{$\Theta_{\ts{e}}\in \left[0,2\pi \right)$}. It is straightforward to show that the eigenvalues of the matrix in the quadratic term of (\ref{EQN09}) are
\begin{align}
\begin{split}
\lambda_1 = 0\quad\ts{and}\quad\lambda_2 = \frac{k_{\delta}^2+1}{2k_{\bs{\alpha}}}
\end{split}.
\end{align}
Therefore, the quadratic term in (\ref{EQN09}) is positive semidefinite and $V$ is positive definite because the integral term is positive definite with respect to \mbox{$\Theta_{\ts{e}}\in \left[0,2\pi \right)$}, for any extended class $\mathcal{K}$ function $\gamma$. To continue, we expand $V$ as \begin{align}
\begin{split}
V &= \frac{k_{\delta}^2}{2k_{\bs{\alpha}}}\gamma^2(\Theta_\ts{e}) + \frac{k_{\delta}}{k_{\bs{\alpha}}}\gamma(\Theta_\ts{e})\bs{u}_\ts{e}^T\bs{\omega}_\ts{e} + \frac{1}{2k_{\bs{\alpha}}}\bs{\omega}_\ts{e}^T\bs{\omega}_\ts{e}\\ 
&\hspace{2ex}+ \int_0^{\Theta_\ts{e}}\gamma(\phi)d\phi,
\end{split}
\label{EQN11}
\end{align}
and compute its time derivative, which yields
\begin{align}
\begin{split}
\dot{V} & = \frac{k_{\delta}^2}{k_{\bs{\alpha}}}\gamma(\Theta_\ts{e})\frac{d\gamma}{d\Theta_\ts{e}}\dot{\Theta}_\ts{e} \\
&\hspace{2ex}+ \frac{k_{\delta}}{k_{\bs{\alpha}}} \left[ \frac{d\gamma}{d\Theta_\ts{e}}\dot{\Theta}_\ts{e}\bs{u}_\ts{e}^T\bs{\omega}_\ts{e}+\gamma(\Theta_\ts{e})\bs{\dot{u}}_\ts{e}^T\bs{\omega}_\ts{e}+\gamma(\Theta_\ts{e})\bs{u}_\ts{e}^T\bs{\dot{\omega}}_\ts{e} \right] \\ 
&\hspace{2ex}+ \frac{1}{k_{\bs{\alpha}}}\bs{\omega}_\ts{e}^T\bs{\dot{\omega}}_\ts{e} + \gamma(\Theta_\ts{e})\dot{\Theta}_\ts{e}.
\end{split}
\label{EQN12}
\end{align}
This expression can be simplified by recalling that \mbox{$\dot{\Theta}_{\ts{e}}= \bs{u}_{\ts{e}}^T \bs{\omega}_{\ts{e}}$}, noticing that \mbox{$\dot{\bs{u}}_{\ts{e}}^T\bs{u}_{\ts{e}}=0$} because $\bs{u}_{\ts{e}}\in \bs{\mathcal{S}}^2$, and substituting (\ref{EQN07b}) into (\ref{EQN12}). Namely,
\begin{align}
\begin{split}
\dot{V} & = \frac{k_{\delta}^2}{k_{\bs{\alpha}}}\gamma(\Theta_\ts{e})\frac{d\gamma}{d\Theta_\ts{e}}\bs{u}_\ts{e}^T\bs{\omega}_\ts{e} \\
&\hspace{2ex}+ \frac{k_{\delta}}{k_{\bs{\alpha}}} \left[ \frac{d\gamma}{d\Theta_\ts{e}}\bs{u}_\ts{e}^T\bs{\omega}_\ts{e}\bs{u}_\ts{e}^T\bs{\omega}_\ts{e}+\gamma(\Theta_\ts{e})\bs{\dot{u}}_\ts{e}^T\bs{\omega}_\ts{e}-k_{\bs{\alpha}}\gamma^2(\Theta_\ts{e})\right.\\
&\hspace{2ex}\left.- k_{\delta}\gamma(\Theta_{\ts{e}})\frac{d\gamma}{d\Theta_\ts{e}}(\bs{u}_\ts{e}^T\bs{\omega}_\ts{e})-k_{\bs{\omega}}\gamma(\Theta_{\ts{e}})\bs{u}_\ts{e}^T\bs{\omega}_\ts{e} \right] \\ 
&\hspace{2ex}+ \frac{1}{k_{\bs{\alpha}}}\bs{\omega}_\ts{e}^T\biggl[-k_{\bs{\alpha}}\gamma(\Theta_\ts{e})\bs{u}_\ts{e} - k_{\delta} \biggl( \gamma(\Theta_\ts{e})\bs{\dot{u}}_\ts{e}\biggr.\biggr.\\
&\hspace{2ex}\left.\left.+\frac{d\gamma}{d\Theta_\ts{e}}(\bs{u}_\ts{e}^T\bs{\omega}_\ts{e})\bs{u}_\ts{e}\right)-k_{\bs{\omega}}\bs{\omega}_\ts{e}\right] + \gamma(\Theta_\ts{e})\bs{u}_\ts{e}^T\bs{\omega}_\ts{e}.
\end{split}
\label{EQN13}
\end{align}
Thus, after canceling out the terms with equal magnitudes and opposite signs, we obtain
\begin{align}
\begin{split}
\dot{V} &= - k_{\delta}\gamma^2(\Theta_\ts{e}) -\frac{k_{\delta} k_{\bs{\omega}}}{k_{\bs{\alpha}}}\gamma(\Theta_\ts{e})\bs{u}_\ts{e}^T\bs{\omega}_\ts{e} - \frac{k_{\bs{\omega}}}{k_{\bs{\alpha}}}\bs{\omega}_\ts{e}^T\bs{\omega}_\ts{e}\\
&= - \begin{bmatrix}
\gamma(\Theta_\ts{e})\bs{u}_\ts{e}^T & \bs{\omega}^T_\ts{e}\\
\end{bmatrix}
\bs{W}
\begin{bmatrix}
\gamma(\Theta_\ts{e})\bs{u}_\ts{e} \\ \bs{\omega}_\ts{e}\\
\end{bmatrix},
\end{split}
\label{EQN14}
\end{align}
where
\begin{align}
\begin{split}
\bs{W} = 
\begin{bmatrix}
k_{\delta} & \dfrac{k_{\delta} k_{\bs{\omega}}}{2k_{\bs{\alpha}}} 
\vspace{1ex}
\\
\dfrac{k_{\delta} k_{\bs{\omega}}}{2k_{\bs{\alpha}}} & \dfrac{k_{\bs{\omega}}}{k_{\bs{\alpha}}}
\end{bmatrix},
\end{split}
\label{EQN15}
\end{align}
which can be made positive definite by selecting
\begin{align}
\begin{split}
\{k_{\delta},k_{\bs{\omega}}\} > 0\quad \ts{and} \quad k_{\bs{\alpha}} > \frac{k_{\delta} k_{\bs{\omega}}}{4},
\end{split}
\label{EQN16}
\end{align}
thus also enforcing that \mbox{$\dot{V}\left(\Theta_\ts{e},\bs{\omega}_\ts{e} \right) <0$} for all $
\{\Theta_\ts{e},\bs{\omega}_\ts{e}\}
\neq
\{\Theta_\ts{e}^{\star},\bs{\omega}_\ts{e}^{\star}\}$, and \mbox{$ \dot{V}\left(\Theta_\ts{e}^{\star},\bs{\omega}_\ts{e}^{\star} \right) = 0$}. In summary, $V$ is positive definite on $\bs{\mathcal{D}}$ and coercive in $\bs{\omega}_{\ts{e}}$; therefore, proper on $\bs{\mathcal{D}}$. Moreover, since \mbox{$\dot{V}$} is negative definite on $\bs{\mathcal{D}}$, we conclude that the unique equilibrium of the CL dynamics given by \mbox{(\ref{EQN07a})--(\ref{EQN07b})} is globally asymptotically stable on $\bs{\mathcal{D}}$, which, together with $\bs{\mathcal{S}}^2$, provides a complete covering of $\bs{\mathcal{SO}}(3)$. \hfill $\square$

\subsection{Discussion}
\label{SEC03D}
There exist infinitely many valid choices of $\gamma$ functions; some superior to the rest. For example, a sigmoid function is an extended class $\mathcal{K}$ function that provides tunable saturation while maintaining linearity near the equilibrium. Namely,  
\begin{align}
\gamma(\Theta_\ts{e}) = \Theta_\ts{max} \frac{1-e^{-\xi\frac{\Theta_\ts{e}}{\Theta_\ts{max}}}}{1+e^{-\xi\frac{\Theta_\ts{e}}{\Theta_\ts{max}}}},
\label{EQN17}
\end{align}
where $\Theta_{\ts{max}}$ and $\xi$ are design parameters. In this context, $\Theta_{\ts{max}}$ is the angle at which the maximum magnitude of the actuation torque occurs and $\xi$ determines the rate at which the magnitude of $\gamma(\Theta_{\ts{e}})$ changes with respect to $\Theta_{\ts{e}}$. Taking the derivative of (\ref{EQN17}) with respect to $\Theta_{\ts{e}}$ yields
\begin{align}
\begin{split}
\frac{d\gamma}{d\Theta_\ts{e}} = \frac{2\xi e^{-\xi\frac{\Theta_\ts{e}}{\Theta_\ts{max}}}}{\left(1+e^{-\xi\frac{\Theta_\ts{e}}{\Theta_\ts{max}}}\right)^2},
\end{split}
\label{EQN18}
\end{align}
which provides insights and can be used to guide the process of controller synthesis. In particular, the slope at the origin has the form \mbox{$\frac{d\gamma}{d\Theta_\ts{e}} (0) = \frac{\xi}{2}$}, where $\xi$ can be tuned considering actuator limitations and control objectives.
\begin{figure}[t!]
\vspace{1ex}
\begin{center}
\includegraphics{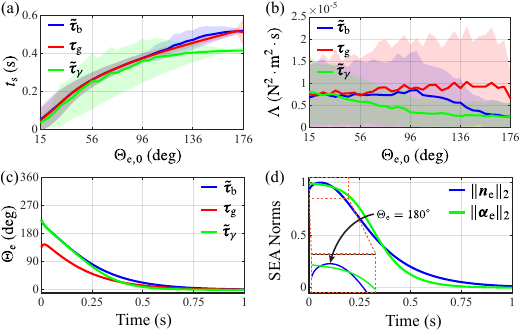}
\end{center}
\vspace{-1ex}
\caption{\textbf{Numerical results obtained with Crazyflie\,2.1 parameters, using quaternion-based, geometric, and axis--angle control laws.} \textbf{(a)}~Mean and ESD of the stabilization time as functions of the initial rotation error. \textbf{(b)}~Mean and ESD of the control effort as functions of the initial rotation error. \textbf{(c)}~Time evolution of the rotation error associated with the \mbox{attitude-error} Euler axis. \textbf{(d)}~Time evolution of the SEA magnitudes corresponding to the simulated \mbox{quaternion-based} and proposed \mbox{axis--angle} laws. \label{Fig02}}
\vspace{-2ex}
\end{figure}

\section{Simulation and Experimental Results} \label{SEC04}
\subsection{Numerical Simulations}
\label{SEC04A}
To systematically assess the functionality and \mbox{performance} of the proposed generalized control law, we compared a controller synthesized using this new approach with two benchmark schemes through numerical \mbox{simulations. The first} { 
\makebox[3.48in][s]{scheme used for comparison uses the \mbox{quaternion-based} atti-}\newline 
tude control law specified \mbox{by (\ref{EQN03}), $\bs{\tau}_{\ts{b}}$; the second is the high-} performance geometric approach presented in\cite{LeeT2011}, whose \makebox[3.48in][s]{corresponding attitude control law here we denote~by~$\bs{\tau}_{\ts{g}}$.}\newline We implemented~and executed the numerical simulations in \mbox{Simulink\,25.2} (\mbox{MATLAB\,R2025b}), using the \mbox{Dormand--Prince} \makebox[3.48in][s]{algorithm with a fixed time step of $10^{-4}\,\ts{s}$. \mbox{The simulations}}\newline \makebox[3.48in][s]{were set up using the \mbox{open-loop} dynamical model speci-}\newline fied by \mbox{(\ref{EQN01a})--\eqref{EQN01b}} with the parameters of the \mbox{Crazyflie\,2.1} \makebox[3.48in][s]{platform---shown in Fig.\,\ref{Fig01}---and controllers with empir-}\newline \makebox[3.48in][s]{ically selected gains that satisfy the stability condi-}\newline \makebox[3.48in][s]{tions for the three compared laws. Specifically, we used} \mbox{$\bs{J} = \ts{diag}\{16.6,16.7,29.3\} \times 10^{-6}\,\ts{kg}\cdot\ts{m}^2$}, \mbox{$k_{\bs{\qbars}} = 10^3\,\ts{s}^{-2}$}, and \makebox[3.48in][s]{\mbox{$k_{\bs{\omega}} = 10^2\,\ts{rad}^{-1} \hspace{-0.2ex}\cdot\ts{s}^{-1}$} for $\bs{\tau}_{\ts{b}}$ specified by (\ref{EQN03}). For $\bs{\tau}_{\gamma}$ speci-}\newline \makebox[3.48in][s]{fied by (\ref{EQN05}), we used the same $\bs{J}$ and $k_{\bs{\omega}}$ that for $\bs{\tau}_{\ts{b}}$,}} and \mbox{$k_{\bs{\alpha}} = 10^3\,\ts{rad}^{-1} \hspace{-0.2ex}\cdot\ts{s}^{-2}$}, \mbox{$k_{\delta} = 10\,\ts{rad}^{-1} \hspace{-0.2ex} \cdot\ts{s}^{-1}$}, and the function $\gamma(\Theta_{\ts{e}})$ defined by (\ref{EQN17}). For $\gamma(\Theta_{\ts{e}})$, we used \mbox{$\Theta_{\ts{max}} = 1$\,rad} and \mbox{$\xi=1.5$}, which were chosen considering the characteristics of the actuators driving the experimental platform. The controller gains for $\bs{\tau}_{\ts{g}}$ were specified to ensure a fair comparison. For consistency with previous published results, we implemented both $\bs{\tau}_{\gamma}$ and $\bs{\tau}_{\ts{b}}$ in combination with the \textit{model predictive selection} (MPS) algorithm presented in\cite{GoncalvesFMFR2024I}, according to which the instantaneous direction of $\bs{u}_{\ts{e}}$, encoded by \mbox{$\sigma \in \{-1,+1\}$}, minimizes

\vspace{-2ex}
{\small
\begin{align}
\Gamma(\sigma, t) = \int^{t+t_\ts{h}}_t \left[ \bs{\tau}^T(\sigma,\zeta)\bs{R}\bs{\tau}(\sigma,\zeta) + \bs{n}_\ts{e}^T(\zeta)\bs{Q}\bs{n}_\ts{e}(\zeta)\, \right] d\zeta,
\label{EQN19}
\end{align}
}

\vspace{-0.5ex}
\noindent in which \mbox{$t_{\ts{h}}=0.2$\,s}; \mbox{$\bs{R} = \bs{I}$}, where $\bs{I}$ is the identity matrix; and, \mbox{$\bs{Q} = 10^{-6} \hspace{-0.2ex} \times \bs{I}~\ts{N}^2 \hspace{-0.2ex} \cdot \ts{m}^2$}. In the simulations, \mbox{$\bs{\tau} \in \{\bs{\tilde{\tau}}_{\ts{b}}(\sigma), \bs{\tilde{\tau}}_{\gamma}(\sigma)\}$}, with $\bs{\tilde{\tau}}_{\ts{b}}(\sigma)$ defined as \mbox{in\cite{GoncalvesFMFR2024I} and}

\vspace{-2ex}
{\small
\begin{align}
\bs{\tilde{\tau}}_{\gamma}(\sigma) = \bs{J}\left[ k_{\bs{\alpha}}\bs{\alpha}_\ts{e}(\sigma) + k_{\delta}\bs{\dot{\alpha}}_\ts{e}(\sigma) + k_{\bs{\omega}} \bs{\omega}_\ts{e} + \bs{\dot{\omega}}_\ts{d}\right] + \bs{\omega}\times\bs{J}\bs{\omega},
\label{EQN20}
\end{align}
}

\vspace{-0.5ex}
\noindent in which \mbox{$\bs{\alpha}_\ts{e}(\sigma) = \sigma\gamma(\Phi_{\ts{e}})\bs{u}_{\ts{e}}$} and \mbox{$\Phi_{\ts{e}} = \left(1-\sigma\right)\pi + \sigma \Theta_{\ts{e}}$}. Note that for \mbox{$\sigma = +1$}, \mbox{$\Phi_{\ts{e}} = \Theta_{\ts{e}}$}, and for \mbox{$\sigma = -1$}, \mbox{$\Phi_{\ts{e}} = 2\pi -\Theta_{\ts{e}}$}.

In total, we performed $10\hspace{0.2ex}908$ simulations of \mbox{high-speed} \mbox{tumble-recovery} maneuvers, in which the control objective was to stabilize the orientation of the UAV from an initial state with arbitrary attitude and angular velocity. Consistent with attitude regulation, for each simulation, \mbox{the desired state was} set to \mbox{$\bs{\qbar}_{\ts{d}} = [1~0~0~0]^T$} and \mbox{$\bs{\omega}_{\ts{d}} = \bs{0}\,\ts{rad}\cdot\ts{s}^{-1}$}. The directions of the initial Euler axes of rotation,
\mbox{$\bs{u}_0 = \bs{u}(t_0)$}, were randomly selected from the unit sphere using a uniform distribution, and the initial rotations about $\bs{u}_0$, \mbox{$\Theta_0 = \Theta(t_0)$}, were taken from the set \mbox{$\left[1\hspace{-0.2ex}:\hspace{-0.2ex}180\right]$\textdegree} in steps of {$5$\textdegree}. For each $\Theta_0$, we varied the initial angular velocity, $\bs{\omega}_0$, from $-30\cdot \bs{u}_0$ to $30\cdot\bs{u}_0\,\ts{rad}\cdot\ts{s}^{-1}$ with the signed magnitude incremented in steps of $0.6\,\ts{rad}\cdot\ts{s}^{-1}$. For each simulated $\Theta_0$ value, we computed the mean and \textit{experimental standard deviation} (ESD) of the set of stabilization times, $\left\{t_{\ts{s}}\right\}$---with $t_{\ts{s}}$ defined as the time it takes for $\Theta_{\ts{e}}$ to reach a value lower than $15$\textdegree---corresponding to the set of $101$ different simulated signed magnitudes of $\bs{\omega}_0$. \mbox{Fig.\,\ref{Fig02}(a)} shows how the mean (solid line) and ESD (\mbox{reduced-opacity} band) of the stabilization time vary as functions of the initial rotation error, \mbox{$\Theta_{\text{e},0}=\Theta_{\text{e}}(t_0)$}, for the three controllers. As seen, the means of the stabilization times obtained with the three controllers are similar for small values of $\Theta_{\ts{e},0}$; however, for larger values of $\Theta_{\ts{e},0}$, the means obtained with $\bs{\tilde{\tau}}_{\gamma}$ are notably smaller, even though $\bs{\tau}_{\ts{g}}$ produces a better performance in some cases---evidenced by the ESD data. 

\mbox{Fig.\,\ref{Fig02}(b)} shows the mean and ESD of the set of control efforts, $\left\{ \Lambda\right\}$---where \mbox{$\Lambda = \int_0^1\|\bs{\tau}(t)\|^2_2dt$}, for \mbox{$\bs{\tau} \in \{\bs{\tilde{\tau}}_{\ts{b}},\bs{\tau}_{\ts{g}},\bs{\tilde{\tau}}_{\gamma}\}$}---as functions of $\Theta_{\ts{e},0}$, for the same simulations corresponding to \mbox{Fig.\,\ref{Fig02}(a)}. Although the means of the control efforts obtained with the three controllers are similar for small values of $\Theta_{\ts{e},0}$, the values of $\Lambda$ corresponding to $\bs{\tilde{\tau}}_{\ts{b}}$ and $\bs{\tilde{\tau}}_{\gamma}$ decrease as $\Theta_{\ts{e},0}$ increases. Notably, the controller designed using the proposed method consistently requires less control effort while achieving a lower mean stabilization time compared to the two benchmark controllers. \mbox{Fig.\,\ref{Fig02}(c)} shows the time evolution of $\Theta_{\ts{e}}$ corresponding to the three simulated control laws, with an initial condition of \mbox{$\Theta_0 = 136$\textdegree} and \mbox{$\bs{\omega}_0 = 30\cdot\bs{u}_0\,\ts{rad}\cdot\ts{s}^{-1}$}. This initial state coincides with a kinematic condition in which the direction of the system's initial angular velocity is opposite to that of the shorter rotational path required to eliminate the attitude error. As seen, the controller based on $\bs{\tilde{\tau}}_{\gamma}$ achieves a stabilization time of \mbox{$0.45$\,s} whereas the benchmark controllers---with laws $\bs{\tilde{\tau}}_{\ts{b}}$ and $\bs{\tau}_{\ts{g}}$---achieve stabilization times of $0.58$ and \mbox{$0.49$\,s}, respectively. The initial rotation error, $\Theta_{\ts{e},0}$, for the $\bs{\tilde{\tau}}_{\ts{b}}$ and $\bs{\tilde{\tau}}_{\gamma}$ cases is \mbox{$224$\textdegree}, which means that these two laws applied the proportional component of the control torque in the direction of the longer rotational path required to eliminate the attitude error. \mbox{Fig.\,\ref{Fig02}(d)} shows the corresponding SEA norms, $\|\bs{n}_{\ts{e}}\|_2$ and $\|\bs{\alpha}_{\ts{e}}\|_2$. As seen, $\|\bs{n}_{\ts{e}}\|_2$ increases as the rotation error, $\Theta_{\ts{e}}$, decreases from \mbox{$224$\textdegree} down to \mbox{$180$\textdegree} and starts decreasing after passing through that point (see inset), which highlights the characteristic behavior of \mbox{quaternion-based} controllers. In contrast, $\|\bs{\alpha}_{\ts{e}}\|_2$ increases and decreases with $\Theta_{\ts{e}}$, as intended by design.
\begin{figure}[t!]
\vspace{1ex}
\begin{center}
\includegraphics{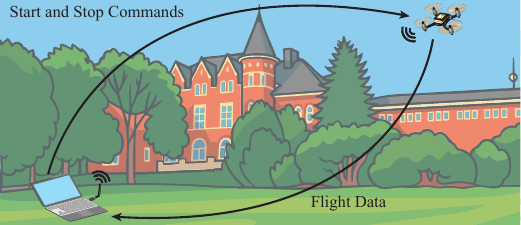}
\end{center}
\vspace{-1ex}
\caption{\textbf{Experimental setup used during the flight tests.} The attitude control flight experiments were run outdoors using a ground computer equipped with the Crazyradio\,2.0 to communicate with the Crazyflie\,2.1. \label{Fig03}}
\vspace{-2ex}
\end{figure}

\subsection{Real-Time Flight Experiments} \label{SEC04B}
To further assess and demonstrate the suitability and performance of the proposed \mbox{axis--angle} control approach, we performed \mbox{real-time} flight experiments using the same three controllers described in \mbox{Section\,\ref{SEC04A}}. As depicted in \mbox{Fig.\,\ref{Fig03}}, the flight tests were conducted outdoors, using a ground computer---equipped with the \mbox{Crazyradio\,2.0} \mbox{$2.4$-GHz} USB radio transceiver---to send initialization and stop commands, and collect the flight and control data at a rate of \mbox{$100$\,Hz}. During each test, the UAV was commanded to execute a \mbox{high-speed} \mbox{tumble-recovery} maneuver after a throw launch. A representative experiment consists of two stages. First, the tested UAV is thrown into the air with an unknown high angular velocity, while the controller and propellers remain inactive; then, the controller and propellers are activated. Throughout a maneuver, the flier operates entirely autonomously, using only onboard sensing and computation, with the tested attitude controller running at \mbox{$500$\,Hz}. 
\begin{figure}[t!]
\vspace{1ex}
\begin{center}
\includegraphics{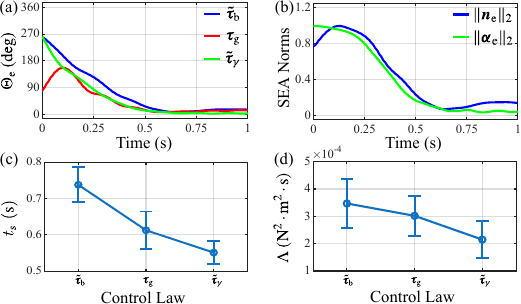}
\end{center}
\vspace{-1ex}
\caption{\textbf{Experimental flight results.} \textbf{(a)}~Rotation errors, as functions of time, corresponding to tests performed using $\bs{\tilde{\tau}}_{\ts{b}}$, $\bs{\tau}_{\ts{g}}$, and $\bs{\tilde{\tau}}_{\gamma}$. \textbf{(b)}~Time evolutions of the SEA magnitudes, $\|\bs{n}_{\ts{e}}\|_2$ and $\|\bs{\alpha}_{\ts{e}}\|_2$, corresponding to the tests performed using $\bs{\tilde{\tau}}_{\ts{b}}$ and $\bs{\tilde{\tau}}_{\gamma}$, respectively, shown in (a). \textbf{(c)}~Comparison of the means (circles) and SEM values (vertical bars) of the stabilization times, corresponding to 30 independent tests per control law, for flights performed using $\bs{\tilde{\tau}}_{\ts{b}}$, $\bs{\tau}_{\ts{g}}$, and $\bs{\tilde{\tau}}_{\gamma}$. \textbf{(d)}~Comparison of the means and SEM values of the control efforts corresponding to the same 30 tests per control law of the data shown in (c). \label{Fig04}}
\vspace{-2ex}
\end{figure}

\mbox{Fig.\,\ref{Fig04}(a)} presents the time evolution of $\Theta_{\ts{e}}$ for three different experiments with similar initial conditions, using the three tested control laws. As seen, the stabilization times corresponding to $\bs{\tilde{\tau}}_{\ts{b}}$, $\bs{\tau}_{\ts{g}}$, and $\bs{\tilde{\tau}}_{\gamma}$ are $0.57$, $0.50$, and \mbox{$0.48$\,s}, respectively. Interestingly, the corresponding values of $\Lambda$ for the $\bs{\tau}_{\ts{g}}$ and $\bs{\tilde{\tau}}_{\gamma}$ laws are \mbox{$1.76\times10^{-4}$} and \mbox{$4.56\times10^{-5}\,\ts{N}^2\hspace{-0.2ex}\cdot\ts{m}^2\hspace{-0.2ex}\cdot\ts{s}$}, which indicates that the proposed method achieved a lower stabilization time using only about \mbox{$25$\hspace{0.2ex}\%} of the control effort measured for the \mbox{geometric-control} case. \mbox{Fig.\,\ref{Fig04}(b)} presents the time evolutions of the SEA norms $\|\bs{n}_{\ts{e}}\|_2$ and $\|\bs{\alpha}_{\ts{e}}\|_2$. Similarly to the simulation cases discussed in \mbox{Section\,\ref{SEC04A}}, it can be observed that $\|\bs{n}_{\ts{e}}\|_2$ increases as $\Theta_{\ts{e}}$ decreases down to \mbox{$180$\textdegree} from approximately \mbox{$270$\textdegree}, and decreases after passing through this point. In contrast, over the entire range of operation, $\|\bs{\alpha}_{\ts{e}}\|_2$ decreases and increases with $\Theta_{\ts{e}}$, as intended by design. Experiments of this type can be viewed in the video available as a Supplemental Item.

In \mbox{Fig.\,\ref{Fig04}(c)}, each data point indicates the mean and \textit{standard error of the mean} (SEM) of the stabilization times corresponding to $30$ \mbox{back-to-back} experiments performed using each of the three control laws compared in this letter. The means of the stabilization times corresponding to $\bs{\tilde{\tau}}_{\ts{b}}$, $\bs{\tau}_{\ts{g}}$, and $\bs{\tilde{\tau}}_{\gamma}$ are $0.74$, $0.61$, and \mbox{$0.55$\,s}, respectively. In \mbox{Fig.\,\ref{Fig04}(d)}, each data point indicates the mean and SEM of the control efforts, $\Lambda$, corresponding to the same experiments of the data shown in \mbox{Fig.\,\ref{Fig04}(c)}. The means of the $\Lambda$ values corresponding to $\bs{\tilde{\tau}}_{\ts{b}}$, $\bs{\tau}_{\ts{g}}$, and $\bs{\tilde{\tau}}_{\gamma}$ are \mbox{$3.47\times10^{-4}$}, \mbox{$3.02\times10^{-4}$}, and \mbox{$2.15\times 10^{-4}\,\ts{N}^2\hspace{-0.2ex}\cdot\ts{m}^2\hspace{-0.2ex}\cdot\ts{s}$}, respectively. Although the differences between these values are not substantial, the recorded experimental data provide compelling evidence of the suitability and superior performance of the proposed approach; particularly, under nonideal conditions due to the presence of actuator saturation, model uncertainty, and disturbances. We infer that these \mbox{real-world} conditions also explain the discrepancies between the simulated and experimental $\Lambda$ values. The large variability in the experimental data is expected due to the random nature of \mbox{tumble-recovery} experiments.

\section{Conclusion} \label{SEC05}
We introduced a new class of \mbox{axis--angle} attitude control laws that provide substantial design flexibility with guaranteed CL stability while ensuring a greater proportional control effort the farther the system's state is from the unique stable fixed AEQ of the CL dynamics. This characteristic is particularly useful when the proposed approach is used in combination with intelligent switching schemes. Evidence regarding functionality and high performance was provided in the form of data obtained through simulations and outdoor flight tests, during which we commanded a quadrotor to autonomously execute \mbox{tumble-recovery} maneuvers in the air. From a statistical standpoint, the new method consistently achieved reduced stabilization times and required less control effort compared to the \mbox{quaternion-based} and \mbox{geometric-control} methods used as benchmarks. 

\bibliographystyle{IEEEtran}
\bibliography{references}

\end{document}